\documentclass[aps,prb,reprint,amsmath,amssymb,superscriptaddress]{revtex4-2}

\usepackage{graphicx}
\usepackage{dcolumn}
\usepackage{hyperref}
\usepackage{bm}
\usepackage[utf8]{inputenc}
\usepackage[T1]{fontenc}
\usepackage{color}

\begin{document}
\title{Quantum versus classical nature of a low-temperature magnetic phase transition in TbAl$_{3}$(BO$_{3}$)$_4$}

\author{T.~Zajarniuk}

\author{A.~Szewczyk}
\email{szewc@ifpan.edu.pl}

\affiliation{Institute of Physics, Polish Academy of Sciences, Al.~Lotników 32/46, PL-02668 Warsaw, Poland}

\author{P.~Wiśniewski}

\affiliation{Institute of Low Temperature and Structure Research, Polish Academy of Sciences, ul.~Okólna 2, PL-50422 Wrocław, Poland}

\author{M.~U.~Gutowska}
\author{R.~Puzniak}
\author{H.~Szymczak}
\affiliation{Institute of Physics, Polish Academy of Sciences, Al.~Lotników 32/46, PL-02668 Warsaw, Poland}

\author{I.~Gudim}

\affiliation{ Kirensky Institute of Physics, Federal Research Center KSC SB RAS, Krasnoyarsk 660036, Russia}

\author{V.~A.~Bedarev}

\affiliation{B. Verkin  Institute for Low Temperature Physics and Engineering of the National Academy of Sciences of Ukraine, 47 Nauky Ave, UA-61103 Kharkiv, Ukraine}

\author{M.~I.~Pashchenko}

\affiliation{B. Verkin  Institute for Low Temperature Physics and Engineering of the National Academy of Sciences of Ukraine, 47 Nauky Ave, UA-61103 Kharkiv, Ukraine}

\affiliation{Institute of Physics, Czech Academy of Sciences, Cukrovarnická 10, 162 00 Praha 6, Czech Republic}

\author{P.~Tomczak}
\affiliation{Faculty of Physics, Adam Mickiewicz University, Uniwersytetu Poznańskiego 2, PL-61614 Poznań, Poland}

\author{W.~Szuszkiewicz}
\affiliation{Institute of Physics, Polish Academy of Sciences, Al.~Lotników 32/46, PL-02668 Warsaw, Poland}

\begin{abstract}
Specific heat, $C_B$, of a TbAl$_{3}$(BO$_{3}$)$_4$ crystal was studied for 50~mK $<T<$ 300 K, with emphasis on $T<1$~K, where a phase transition was found at $T_c =0.68$~K. Nuclear, non-phonon ($C_m$), and lattice contributions to $C_B$ were separated. Lowering of $T_c$ with ncrease of magnetic field parallel to the easy magnetization axis, $B_{||}$, was found. It was established that $C_m$ and a Gr\"uneisen ratio depend on $B_{||}$ and $T$ in a way characteristic of systems, in which a classical transition is driven by quantum fluctuations, QF, to a quantum critical point at $T=0$, by tuning a control parameter ($B_{||}$). The $B_{||} - T$ phase diagram was constructed and the dynamical critical exponent $0.82 \le z \le 0.96$ was assessed.
Nature of the transition was not established explicitly. Magnetization studies point at the ferromagnetic ordering of Tb$^{3+}$ magnetic moments, however, lowering of $T_c$ with increase in $B_{||}$ is opposite to the classical behavior. Hence, a dominant role of QF was  supposed.
\end{abstract}

\maketitle

\section{Introduction}
Quantum phase transitions, QPT, induced at zero temperature, $T$, by quantum fluctuations, QF, as the result of tuning a certain control parameter, e.g., pressure or magnetic field, $B$, are a topical subject of research in condensed matter physics \cite{sach}. As the phenomena appearing at inaccessible experimentally $T=0$, they are much more difficult for investigation than the classical phase transitions, induced by thermal fluctuations. Their existence can be recognized only by investigating certain unusual properties induced by them at finite $T$ near the quantum critical point, QCP. For example, near QCP, unconventional superconductivity and pronounced non-Fermi-liquid effects in metallic systems were observed \cite{gegen}. Moreover, at the classical transitions, some thermodynamic quantities being the second derivatives of the thermodynamic potential diverge, whereas due to the third law of thermodynamics, some of these divergences, e.g., of specific heat, disappear at QCP. According to \cite{zhu} and \cite{garst}, in such a case, the parameter $\Gamma$:
\begin{equation}
\Gamma= -\frac{1}{T} \frac{({\partial S}/{\partial B})_T}{({\partial S}/{\partial T})_B}=-\frac{({\partial M}/{\partial T})_B}{C_{B}\left (T\right )}=\frac{1}{T}\left ( \frac{\partial T}{\partial B}\right)_S
\label{eq1}
\end{equation}
called the magnetic Gr\"uneisen ratio, is much more informative, because it should diverge and change sign at QCP. $\Gamma$ is proportional to the ratio of the sensitivity of entropy, $S$, to the control parameter $B$ (i.e., $\partial S/ \partial B$), growing near QPT, to the sensitivity of $S$ to $T$ (i.e., $\partial S/ \partial T$), growing near the classical transition. By using standard thermodynamic transformations and the Maxwell relation, it can be shown (\ref{eq1}) that $\Gamma$ is the ratio of the measurable quantities, i.e., of the minus derivative of magnetization, $M,$ with respect to $T$, at fixed $B$, to the specific heat, $C_B$, or as the (multiplied by $1/T$) adiabatic change of $T$ under influence of $B$. The latter value is the main parameter characterizing magnetocaloric effect.

While many papers considered QPT in metallic, heavy-fermion and/or superconducting systems \cite{gegen}, experimental papers on QPT in insulating anti- or ferromagnets are scarce and, in majority, consider pyrochlores \cite{kadowaki} or organic systems, in which magnetic moments  arranged in chains or planes are considered by using the Ising or Heisenberg models, e.g., \cite{ryl}, \cite{gali}, and \cite{wolf}.
\begin{figure*}
\includegraphics[width=1\textwidth]{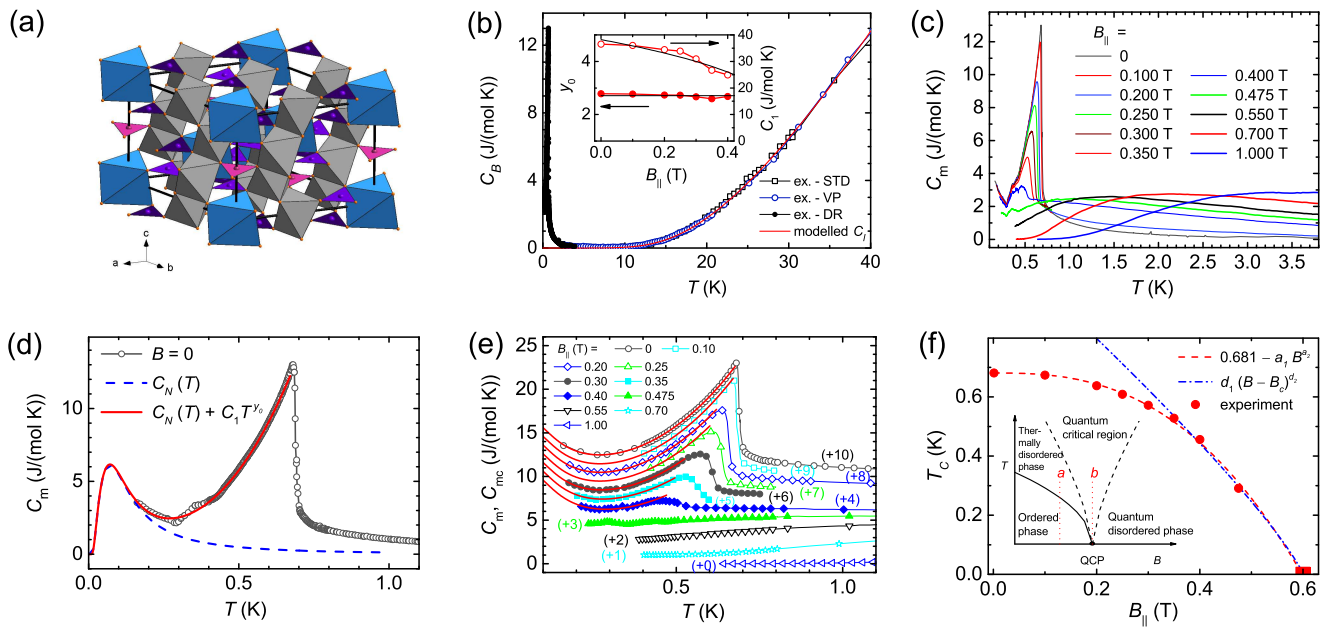}
\caption{\label{fig1}\  Specific heat of the TbAl$_{3}$(BO$_{3}$)$_4$ single crystal. (a) Trigonal structure of TbAl$_{3}$(BO$_{3}$)$_4$ with lattice parameters $a$ = 9.2926(9) \AA, and $c$ = 7.2516(4) \AA . Magnetic Tb$^{3+}$ ions are located inside the deformed (blue) trigonal prisms, located along the 3-fold $c$ axis and formed by six O$^{2-}$ ions.
(b)~Total specific heat, $C_B$, measured in $B=0$ with standard calorimeter, STD, vertical puck, VP, and dilution refrigerator, DR. The red solid line presents the estimated $C_l$. Inset, the $C_1$ and $y_0$ coefficients of the curves, Eq. (\ref{eq5}), fitted to the experimental data presented in (e), as functions of $B_{||}$.
(c)~Non-phonon contribution, $C_m$, to $C_B$ for several $B_{||}$ values.
(d)~Experimental $C_m( T , B=0)$ function (circles), estimated nuclear specific heat, $C_N$ (blue dashed line), and fit of the  $C_N(T) +C_1 T^{y_0}$  function to $C_m( T , B=0)$ (red solid line). Near 330~mK a “sinusoidal” apparatus effect is visible.
(e)~Determined $C_{\mathrm {mc}}(T)$, i.e.\ $C_m(T)$ corrected for the apparatus effect, for $B_{||} \leq 0.475\,$T, and $C_m(T)$ for $B_{||}>0.475\,$T (symbols), and $C_m(T) = C_N(T) +C_1 T^{y_0}$ functions fitted to these curves (red solid lines). To maintain readability, the curves for different $B_{||}$ are shifted along the vertical axis by the values given in parentheses.
(f)~$B_{||}-T$ phase diagram found based on $C_m(T,B_{||})$ functions (symbols). Inset, schematic phase diagram, in which a line of classical phase transition ends at QCP.
}
\end{figure*}
While the quantum Ising system located in $B$ perpendicular to the Ising axis is the basic and intuitively clear example of the system showing QPT, the bulk TbAl$_{3}$(BO$_{3}$)$_4$ crystal seems to be a unique ferromagnetic Ising system, in which the temperature of the magnetic transition, discovered by us below 1~K and reported initially in Ref. \cite{bedarev}, is lowered and driven to, as we suppose, QCP by $B$ applied along the Ising axis. This is a counterintuitive behavior, because in ``normal'' ferromagnets, $B$ applied along the easy axis supports the low temperature phase and lifts the transition temperature. Our paper is aimed (i)~at presenting measurements of $C_B$ and $M$ of the TbAl$_{3}$(BO$_{3}$)$_4$ single crystal, demonstrating the presence and evolution of the magnetic transition under influence of $B$,  (ii)~at analysing whether near the transition point, the specific heat, magnetization, and $\Gamma$ parameter behaviors as functions of $T$ and $B_{||}$ are consistent with universal behaviors, independent of a physical mechanism of the transition, predicted by the renormalization group theory for the second order phase transitions,  (iii)~at analyzing if these behaviors are characteristic of the transitions having a quantum character, i.e., influenced by QF, and (iv)~at considering possible mechanisms of the transition found.

It should be noted that $RT_{3}$(BO$_{3}$)$_4$ crystals, with $R = $~Y or rare earth ion and $T = $~Al, Ga, Cr or Fe, attract attention, because they are suitable for laser applications (e.g., aluminoborates doped with Nd are used in self-doubling frequency lasers), show large magnetoelectric effect \cite{liang}, \cite{zhang} and different magnetocrystalline anisotropy for various $R$ ions \cite{liang2}. Thus, deep knowledge of their properties over a wide temperature range is highly desirable.

\section{Experiment}

$R$Al$_{3}$(BO$_{3}$)$_4$ compounds with $R =$~Y, Sm -- Yb crystallize in a trigonal structure (space group no.\ 155, R32), Fig.~\ref{fig1}a, with three formula units in the trigonal unit cell \cite {leon}. Magnetic $R^{3+}$ ions are located inside the deformed trigonal prisms formed by six O$^{2-}$ ions. The $R$--O$_6$ prisms, separated along the \textit{c} axis by B--O$_3$ triangles, form chains along the 3-fold $c$ axis. Between the neighboring chains, other B--O$_3$ triangles and oxygen octahedra containing Al$^{3+}$ ions are located.

For the 0.514(2) mg plate, cut perpendicularly to the \textit{c} axis from the TbAl$_{3}$(BO$_{3}$)$_4$ single crystal grown by using the flux method \cite{eremin}, \cite{gudim}, specific heat and magnetization were measured as a function of $T$ and magnetic field applied along, $\mathbf B_{||}$, and perpendicularly, $\mathbf B_{\bot}$, to the $c$ axis.

$C_B$ was measured by means of the relaxation method, by using the Quantum Design PPMS system, equipped with the Dilution Refrigerator and Heat Capacity options. To gather the data for estimating the lattice specific heat, $C_l$, the measurements were done from 50~mK to 300~K for $B=0$. Since these studies  showed that a certain unknown phase transition appears at 0.68 K, the detailed $C_B$ studies were performed for the 50~mK -- 4~K range, for $B=0$ and for several $B_{||}$ and $B_{\bot}$ values, up to 3~T. It was verified that for both zero and non-zero $B$, no $C_B$ singularities appear above 4~K. Near the discovered transition, $C_B$ was measured each 2~mK, whereas outside this region, over the range 50~mK -- 1~K, each $\sim 10$~mK.

By using the Quantum Design MPMS - XL SQUID  magnetometer equipped with the Helium 3 option, $M(T)$ functions were measured for several $B_{||}$ values (0.01, 0.1, 0.2, 0.25, and 0.35~T), for 0.5~K~$ \leq~T \leq$~1~K. At $T = $ 0.5~K, the $M(B)$ function was measured for $B_{||}$ and $B_{\bot}$.

\section{Results}
Results of the $C_B$ and $M$ measurements are presented in Figs.~\ref{fig1}~and~\ref{fig2}.

In order to extract the most interesting for the present studies, ``magnetic'' or ``non-phonon'' contribution, $C_m$, to $C_B$, we modelled $C_l(T)$ with the expression:
\begin{equation}
\begin{split}
&C_l\left (T\right )= \Bigg [ 3n_{D} \left ( \frac{T}{\theta_{D}}\right )^3
\int_{0}^{{\theta_D}/{T}} \frac{x^4e^x}{(e^x - 1)^2} dx \\
&+ \sum_{i = 1}^{n_{O}} n_{i} \left ( \frac{\theta_{ i}}{T} \right ) ^{2} \frac{e^{{\theta_i}/{T}}}{(e^{{\theta_{i}}/{T}} - 1)^2} \Bigg ] \frac{k_B N_ A}{(1 - \alpha T)},
\label{eq2}
\end{split}
\end{equation}
\noindent which mimics the contribution related to acoustic and some optical phonons in frames of the Debye model (the first therm in the parentheses), the contribution related to remaining optical phonons in frames of the Einstein model (the second term in the parentheses), and takes into account the effect of thermal lattice expansion by the method proposed in \cite{mart} (the $1-\alpha T$ denominator). It was used successfully for describing the lattice specific heat of many materials, e.g.\ of layered cobaltites  \cite{wiec} and olivines  \cite{lew}. We fitted \eqref{eq2} to the data measured for TbAl$_{3}$(BO$_{3}$)$_4$ in $B=0$ above 15 K, because, as Fig.~\ref{fig1}b illustrates, for $T>15$~K all other than $C_l$ contributions can be neglected. The best description of $C_l(T)$ was achieved by taking the following values of the fitted parameters: $\alpha=$ 0.00129~K$^{-1}$, $n_{D}= 3$, $\theta_{D}= 465$~K, $n_O=6$, and, for $i=$ 1, 2,..,6, respectively, $n_{i}=$ 1, 2, 3, 2, 6, and 6, and $\theta_{ i} =$ 105, 160, 302, 462, 496, and 565~K. The $\theta_{i}$ values represent energies of the lowest optical branches, expressed in temperature units. They agree qualitatively with the energies of optical phonons corresponding to translations of the $R^{3+}$ ions, as well as to translations and librations of the BO${_3}$ complexes, given in \cite{dob}.

\indent $C_m$ determined as the difference between the measured $C_B(T)$ and the estimated $C_l(T)$ is presented in Figs.~\ref{fig1}c and \ref{fig1}d. At low $T$, for $B_{||} < 0.475$~T, the $C_m(T, B_{||}= \text {const})$ functions show: (i)~a $\lambda$-shaped anomaly (at $T=$ 680(2)~mK for $ B=0$), which decreases and shifts towards lower temperatures with increasing $B_{||}$, (ii)~a practically field-independent minimum $C_m =$ 1.10(5)~J/(mol K) at $T_m =$ 290(1)~mK, and (iii) upturn with decreasing $T$ below $T_m$. The temperature range in which this upturn could be studied in our apparatus, was limited by a strong increase of the heat relaxation time with lowering $T$, and as the result, no reliable specific heat values could be measured below 160~mK for $B=0$ and 200~mK for $B_{||}=0.4$~T. For $B_{||}> 0.475$~T, Figs.~\ref{fig1}c and \ref{fig1}e, sudden disappearance of the $\lambda$ anomaly and of the minimum at $T_m$ are observed, and $C_m$ decreases with decreasing $T$ monotonically, down to unmeasurable values. Thus, the ``lowest'' experimental points were measured at $\sim39$~mK for $B_{||}=0.7$~T and at $\sim 640$~mK for $B_{||}=1$~T.

\section{Analysis}
The asymmetric $\lambda$ shape of the anomalies found in the $C_m (T, B=\text {const})$ functions, their rather large width, and lack of  thermal hysteresis of their appearance on heating and on cooling the sample allow to assume that they are related to a second order phase transition. Of  course,  the lack of hysteresis is the necessary, but not sufficient, condition only and the two former arguments are rather heuristic. Thus, a more detailed analysis whether the specific heat as a function of temperature and magnetic field shows universal critical behaviors characteristic of the second order transitions, predicted by the renormalization group theory and independent of a physical mechanism of the transition, is necessary to identify the order of the found transition unequivocally. Such analysis was performed and will be presented below.

\begin{figure}
\includegraphics[width=0.43\textwidth]{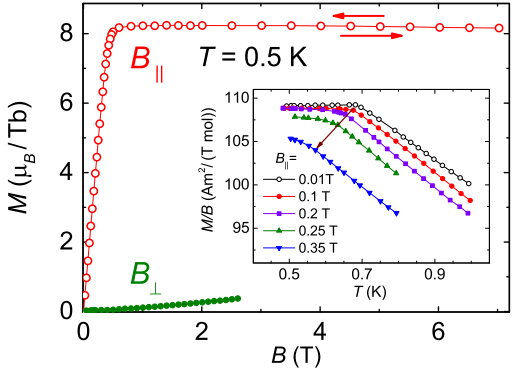}
\caption{\label{fig2}\ Magnetization of the TbAl$_{3}$(BO$_{3}$)$_4$ single crystal (not corrected for demagnetizing effects). Main panel, $M$ versus $B_{||}$ and $B_{\bot}$ at $T=0.5$~K. Inset, $M/B$ versus $T$ for $B_{||}=$ 0.01, 0.1, 0.2, 0.25, and 0.35 T. The arrow indicates evolution of the transition temperature with increase of $B_{||}$.}
\end{figure}

The specific heat data, Fig.~\ref{fig1}, the magnetization curves presented in Fig.~\ref{fig2}, as well as high sensitivity of the phase transition temperature to $B_{||}$ suggest the transition to be related to ordering of magnetic moments of the Tb$^{3+}$ ions, being the only magnetic ions in the system. The shape of the magnetization curve for $B_{||}$, presented in Fig.~\ref{fig2}, and the measured saturation magnetization of TbAl$_{3}$(BO$_{3}$)$_4$, 8.2 $\mu_B$/Tb, being close to the magnetic moment of the free Tb$^{3+}$ (9 $\mu_B$) ion, suggest that we deal with the ferromagnetic ordering. However, if it were the classical transition between the paramagnetic and ferromagnetic phases, induced by thermal fluctuations, $\mathbf B_{||}$ should smear it and shift it towards higher $T$. Actually, we observe the opposite, counterintuitive effect. The $\lambda$ anomaly remains sharp and shifts towards lower $T$ with increase in $B_{||}$, i.e., it behaves in a way characteristic of antiferromagnets. Thus, we suppose that the transition found has a ``quantum'' character, i.e.\ it is dominated by QF, which destroy the long range ferromagnetic order. In other words, we deal with one of the two model cases considered in the physics of quantum transitions  \cite{sach}, \cite{voj}, shown schematically in the inset to Fig.~\ref{fig1}f, in which the line of classical transitions ends at QCP. Below, we analyse, if other characteristics of the transition support this idea. However, the possibility that Tb--Tb exchange interactions are not the main driving force leading to the ferromagnetic order but some other mechanism is responsible for the transition evolving to QCP and the ordering of Tb$^{3+}$ moments is a side effect only, can not be excluded  \emph {a~priori}. An ordering of electric quadrupolar moments of the Tb--O$_6$ complexes, cooperative-Jahn-Teller effect, or a more complex multipolar ordering \cite{santini}, \cite{sivardiere} could be indicated as such possible mechanisms.

To interpret the presence of the minimum at $T_m$ and of the upturn of $C_m (T,\, B=\text {const})$ on lowering $T$ below $T_m$, for $B_{||}<0.475$~T, we assumed $C_m$ to be the sum of the critical contribution, $C_\mathrm {cr}$, related to the phase transition, and the nuclear specific heat, $C_N$, related to excitations of magnetic Tb nuclei. $C_N$ grows on lowering temperature, and becomes a dominating contribution to the measured specific heat, $C_B$, below $T_m$. Following analysis presented in \cite{loun}, we verified that for TbAl$_{3}$(BO$_{3}$)$_4$, the nuclear quadrupole interaction can be neglected. Then, the Hamiltonian of the nucleus can be presented in the form $\hat{H}=a \hat{I}_z$, where $\hat{I}_z$ is the operator of the $z$-component of the nuclear spin and $a$ is the coefficient of the interaction between the nuclear magnetic moment and the hyperfine field, $B_\mathrm {hyp}$. For Tb$^{3+}$,  $I =3/2$ and the nuclear magnetic moment is equal to 1.994~$\mu_N$ ($\mu_N$ is the nuclear magneton). Thus, expressing energy in temperature units, we obtain the following formula for the nuclear contribution to the molar specific heat of TbAl$_{3}$(BO$_{3}$)$_4$:
\begin{equation}
C_N(T)=\frac {R}{T^2}\, \frac {a^2\left[\cosh \left ( {2 a}/{T}\right) + 4 \cosh \left ( {a}/{T} \right) + 5 \right]}{2\,  \left[ \cosh \left(  {3 a}/{2 T} \right) + \cosh \left( {a}/{2 T} \right) \right]^2 },
\label{eq3}
\end{equation}
where $|a|= B_\mathrm {hyp}\cdot1.994\, \mu_N\cdot 2/(3k_B )$, and $R$ denotes the gas constant. In the further analysis, we treated the hyperfine field, $B_\mathrm {hyp}$, as the fitted parameter.

According to \cite{zhu} and \cite{garst}, in the vicinity of a quantum second order transition, the scaling relation for the critical contribution to free energy, $F_\mathrm{cr}$, has a form:
\begin{equation}
F_\mathrm{cr}(T)=-R\, \rho_0 \left | r \right |^{\nu (d+z)} \widetilde f \left ( \frac {T}{T_0\, \left | r \right |^{\nu z}} \right ) ,
\label{eq4}
\end{equation}
where $\widetilde f (x)$ is a universal scaling function, $d$ is dimension of the considered system, $r=(B-B_c)/B_c$ is the control parameter describing the distance from the QCP at $T=0$ and $B=B_c$, $\rho_0$ and $T_0$ are non universal parameters, $\nu$ is the exponent of the critical behavior of the correlation length near QCP, $\xi \sim |r|^{-\nu}$, and  $z$ is the ``dynamical critical exponent'' describing the correlation length $\xi_{\tau} \sim \xi^z$ along the direction of ``imaginary time'' $\tau = i\hslash/ (k_B T)$. For  $B=B_c$, for the system evolving along the trajectory denoted as ``$b$'' in the schematic phase diagram in the inset to Fig.~\ref{fig1}f, the free energy is a regular function of $T$ that can be expanded into the Taylor series. However, at $B<B_c$, for the system evolving along the ``$a$'' trajectory and meeting on its ``way'' the classical phase transition, the free energy shows a more peculiar behavior. In the limit of $T \to 0$ and $r \ne 0 $, $\widetilde f (x)$ can be approximated by the formula $\widetilde f (x \to 0)= \widetilde f (0) + c\, x^{y_0 +1}$, where $y_0$ is a positive exponent. Thus, differentiating (\ref{eq4}) with respect to $T$, one obtains the entropy and next, by differentiating it with respect to $T$ and $B$ one obtains the following formulae for the critical specific heat, $C_\mathrm{cr}$, and $\Gamma$, valid for the area below the line of classical phase transitions, denoted in the inset to Fig.~\ref{fig1}f as ``Ordered phase'':
\begin{equation}
\begin{split}
&C_\mathrm{cr}(T \to 0, B)= T \frac {\partial S} {\partial T} = C_1(B)\, T^{y_0},  \\
&C_1(B) = R\: \frac {\rho_0\, c\, y_0\, (y_0 +1)} {{T_0}^{y_0 +1}} \left | \frac {B-B_c} {B_c} \right | ^{\nu\, (d-y_0z)},
\label{eq5}
\end{split}
\end{equation}
\begin{equation}
\Gamma (T \to 0, B) = - \frac {G_B}{B-B_c}, \, \text {with } \, G_B = \frac {\nu\,(d - y_o \,z)}{y_0},
\label{eq6}
\end{equation}

By assuming $C_m$ to be the sum of $C_N$ and $C_\mathrm{cr}$ (\ref{eq5}):
\begin{equation}
C_m(T, B=\text {const}) = C_N(T)+C_1(B)\,T^{y_0},
\label{eq7}
\end{equation}
and treating $B_\mathrm {hyp}$, $C_1$ and $y_0$ as fitted parameters, a good description of the experimental $C_m(T)$ dependence for $B=0$ (thick solid red line in Fig.~\ref{fig1}d) was achieved. The best fit value $B_\mathrm {hyp}=227.7(1)$~T seems to be reasonable, because for the magnetically ordered terbium metal \cite{heltemes} and alloys \cite{vuayaraghavan}, the values $360 \pm 40$~T were reported. As Fig.~\ref{fig1}c shows, $B_{||} < 0.4$~T has no influence on $C_N$, because for these $B_{||}$ values, all $C_m(T)$ curves overlap below $T_m$. Thus, it was assumed that $B_\mathrm {hyp}$ remains constant for $B_{||} < 0.4$~T, then for larger $B_{||}$, it falls down due to destruction of the long range order of Tb$^{3+}$ magnetic moments, and for $B_{||}>0.55$~T, $C_N$ becomes unmeasurably small. We found that the ``sinusoidal'' anomaly of the experimental $C_m(T, B=0)$ curve, visible in Fig.~\ref{fig1}d between 0.25 and 0.35 K, is not a physical effect, because it is the same for all $B_{||} \le 0.475$~T and appears also for samples of other composition. Thus, it was interpreted as an apparatus effect and eliminated from the experimental $C_m(T,B_{||}=\text {const})$ curves for $B_{||} \le 0.475$~T by subtracting from them the difference between the experimental and theoretical (red solid line in Fig.~\ref{fig1}d) $C_m(T,B=0)$ curves for the range $242 \le T \le 479$~mK. The corrected functions (for $B_{||} \le 0.4$~T) denoted as $C_\mathrm {mc}$ are plotted in Fig.~\ref{fig1}e together with the curves (\ref{eq7}) fitted to them (red solid lines) by taking $B_\mathrm {hyp}=227.7(1)$~T and fitting $\ C_1$ and $y_0$. Next, assuming the $\lambda$ maxima of the $C_m(T, B)$ functions to correspond to the phase transition temperature, we constructed the $B_{||}$--$T$ phase diagram, Fig.~\ref{fig1}f. We found that the power function: 0.681~$\mathrm{K} - a_1 B^{a_2}$, with $a_1=2.65(5)~\mathrm {K}$, $a_2=2.65(5)$, and $B$ given in T describes the experimental phase transition line very well and gives the hypothetical quantum critical field value $B_c=0.600(1)$~T. With this $B_c$ value, the phase transition line for $0.35~\mathrm {T} \le B_{||} \le B_c $ can be described by the typical critical dependence: $d_1  (B_c - B_{||}) ^{d_2} $, with $d_1=1.75(3)$~K, $d_2=0.85(1)$, and $B_{||}$ given in T. Using this $B_c$ in the definition of $r$, we obtained that the $0 \le B_{||} \le 0.475\;\mathrm{T}$ values correspond to $1 \ge r \ge 0.21$. This result is essential, because usually one assumes the critical behavior to appear for $r < 10^{-3} $, thus, the behaviors observed in our experiments should be extrapolated to smaller $r$ values to get actual critical behaviors of the studied system.
\begin{figure}
\includegraphics[width=0.48\textwidth]{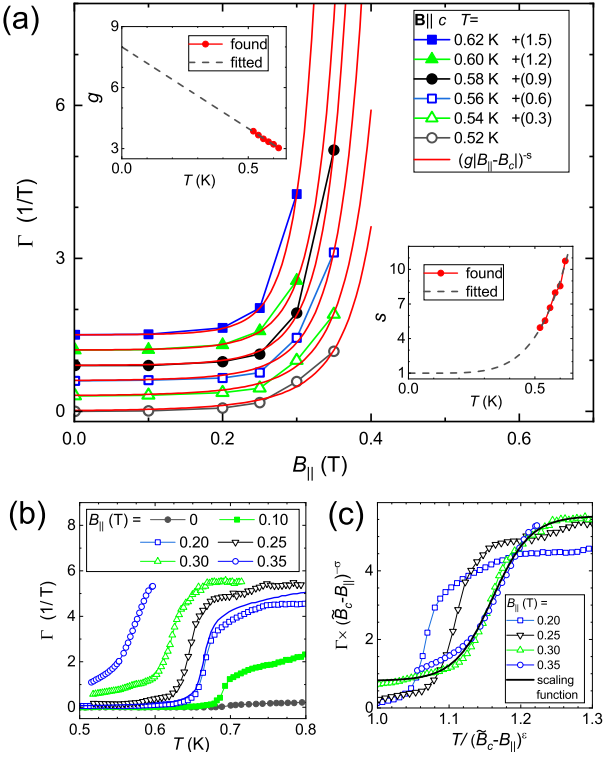}
\caption{\label{fig3}\ The Gr\"uneisen ratio, $\Gamma$. (a) $\Gamma$ values as a function of $B_{||}$, symbols, and the theoretical $g\left |B-B_c \right |^{-s}$ functions fitted to them, solid red lines, for several $T$ values. Curves for different $T$ are shifted along the $\Gamma$ axis by the values given in the legend in parentheses. Insets, $g$ and $s$ parameters as a function of $T$.
(b) $\Gamma(T)$ for several $B_{||}$, symbols. Blue solid line is the $\Gamma (T, B_{||}=0.2\; \mathrm {T})$ function calculated based on the $C_\mathrm{mc} (T)$ curves.
(c) Scaling behavior. $\Gamma \cdot (\widetilde B_c -B_{||})^{-\sigma}$ as a function of $T/(\widetilde B_c -B_{||})^\epsilon$ for $\epsilon = 0.48(4)$, $\sigma = 0 $, and $\widetilde B_c = 0.57(3)$~T, for several $B_{||}$. Black solid line is the approximation of the scaling function: $\Phi(x)=3.2+2.4\, \tanh (19\,(x-1.161))$. The curves for $B_{||} \ge 0.3$~T overlap for the arguments 1.132 -- 1.221.}
\end{figure}

Plots of the found $y_0$ and $C_1$ values as a function of $B_{||}$ (inset to Fig.~\ref{fig1}b) can be approximated by, drawn with the black solid lines, the field independent $y_0=2.71(7)$ value and the $C_1(B_{||})=c_1\left ( B_c - B_{||} \right )^{c_2}$ function with $c_1=45.7$~J/(mol~K), $c_2=0.35$, and $B_{||}$ given in T.  Comparing the $C_1(B_{||})$ function and (\ref{eq5}) we get:
\begin{equation}
z= \left ( d- c_2/\nu \right )/y_0.
\label{eq8}
\end{equation}

The $\Gamma(T)$ functions for different $B_{||}$ were estimated by substituting into (\ref{eq1}) the $C_\mathrm {mc} (T, B_{||})$ functions determined and the $(\partial M/\partial T)_{B_{||}}$ derivatives, approximated by the difference quotients, calculated basing on the measured $M(T ,B_{||})$ functions, Fig.~\ref{fig2}. For $B_{||}=0.3$~T, for which $M$ was not measured, $\partial M/\partial T$ was estimated as the average of the values calculated for $B_{||}=0.25$ and 0.35~T. The results are shown in Fig.~\ref{fig3}b. In principle, having $C_\mathrm {mc}(T)$ for different $B_{||}$, it was possible to calculate the entropies $S(T, B_{||}= \text{const})= \int C_\mathrm{mc}/T\, dT$ and $\partial S/\partial B$. However, $C_B(T)$ was measured for a few $B_{||}$ values only and hence, the approximation of $\partial S/\partial B$ with the difference quotients was rough. Thus, this method of determining $\Gamma$ was less accurate and only one of such determined functions, for $B_{||}=0.2$~T, is shown in Fig.~\ref{fig3}b as an example. The $\Gamma(T)$ functions were converted into $\Gamma(B_{||})$ functions, shown in Fig.~\ref{fig3}a for several fixed $T$ values. It was found that they can not be described by the formula (\ref{eq6}) but they can be approximated by formula:
\begin{equation}
\Gamma (B_{||}, T= \text {const}) = \left [ g \left ( B_c - B_{||} \right ) \right ]^{-s}.
\label{eq9}
\end{equation}
The functions (\ref{eq9}) fitted to the experimental data for different $T$ are plotted in Fig.~\ref{fig3}a with red solid lines and the $g$ and $s$ parameters of the best fit lines are plotted as a function of $T$ in the insets to Fig.~\ref{fig3}a. As the dashed lines show, (i) the $s(T)$ function can be approximated by: $s(T)=1+111\; T^{5.1}$, for $T$ given in K, which, when extrapolated to  $T<0.2$~K, gives the $s(T) \approx 1$ value, consistent with the theoretical prediction (\ref{eq6}), and (ii) $g(T)$ can be approximated by the straight line: $g(T)=8-8\;T$, for T given in K, which allows to assume the parameter $G_B$ appearing in (\ref{eq6}) to be $G_B \approx 1/g(0)=0.125(2)$. Thus, based on (\ref{eq8}) and (\ref{eq6}), using the determined $y_0$, $c_2$, and $g(0)$ parameters, and taking into account experimental uncertainties, we obtain: $1/g(0)=c_2/y_0 \approx 0.126(3)$ and:
\begin{equation}
z= d / y_0 - 1/ [ \nu \: g(0)] = d/y_0 - 0.126/\nu.
\label{eq10}
\end{equation}
By assuming that the we deal with the 3D magnetic system ($d=3$) and that $\nu$ takes a value between 1/2 (found in the molecular field model) and 0.715 (found for the 3D Heisenberg model) we find the dynamical critical exponent of the studied system to be  $0.82 \le z \le 0.96$.

Seeking for a based on scaling argument for the quantum character of the transition analyzed, we based on the observation \cite{gegenwart},  that for some systems, near QCP with the critical field $\widetilde B_c$, $\Gamma$ scales with $T(\widetilde B_c-B )^{-\epsilon}$, with a constant $\epsilon$. Thus, we assumed that at quantum criticality, $\Gamma$  is a generalized homogeneous function relative  $\widetilde B_c-B$, which fulfills the relation
$\Gamma (T,\widetilde B_c-B)\sim (\widetilde B_c-B)^\sigma \Phi \bigl [T(\widetilde B_c-B)^{-\epsilon}\bigr ]$ with a constant $\sigma$ and  $\Phi$ being a scaling function. If the scaling occurs, it should be possible to find such  $\widetilde B_c$, $\sigma$, and $\epsilon$
that in the critical region, the plots $\Gamma  \cdot (\widetilde B_c-B)^{-\sigma}$ versus $T(\widetilde B_c-B)^{-\epsilon}$ for different $B$ values collapse onto a single curve, $\Phi$. As Fig.~\ref{fig3}c shows, such a collapse of the curves for $B_{||} = 0.30$ and 0.35~T was achieved by taking $\widetilde B_c=0.57(3)$~T,  $\epsilon =0.48(4) $ and $\sigma=(437 \pm 2)\, 10^{-6} \approx 0 $. Practically zero $\sigma$ value confirms the $\Gamma$ scaling assumed in \cite{gegenwart}. The data sets for $B_{||} < 0.3$~T do not follow the scaling curve, which suggests that these field values are below the quantum critical region. The $\widetilde B_c=0.57(3)$~T value found from scaling is slightly smaller than that found from extrapolation of the phase transition line shown in Fig.~\ref{fig1}f, which suggests that the real QCP is located between 0.57 an 0.6~T.

\begin{figure}
\includegraphics[width=0.48\textwidth]{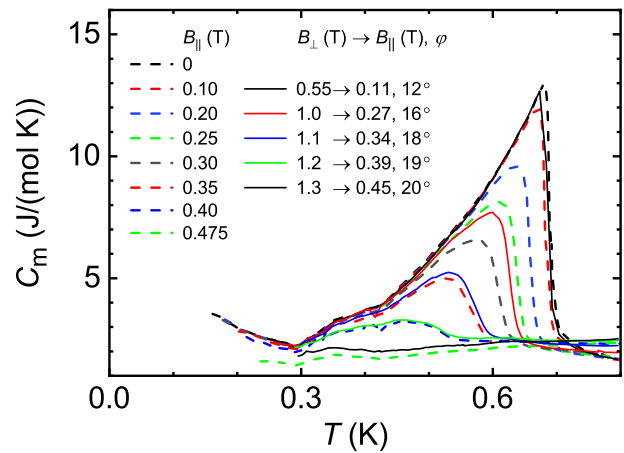}
\caption{\label{fig4}\ $C_m(T)$ functions for several fixed $B_{||}$ (dashed curves) and $B_\bot$ (solid lines), applied intentionally perpendicularly to the $c$ axis, values. In the legend, the expected sample tilting angles, calculated under assumption that only the $B_{||}$ component influences the transition point, as well as corresponding to them $B_{||}$ components are given.}
\end{figure}

In order to study influence of the magnetic field directed perpendicularly to the easy magnetization $c$ axis, $\mathrm B_{\bot}$, on the observed magnetic phase transition, temperature dependence of the specific heat, $C_B$, was measured for several fixed values of the field applied intentionally perpendicularly to the $c$ axis. Next, the non-phonon contribution to it, $C_m$, was determined by subtracting $C_l$ from the measured $C_B$, i.e., in the same way as for the case of $\mathrm B_{||}$. As Fig.~\ref{fig4} shows, $\mathrm B_{\bot} $, applied intentionally perpendicularly to the $c$ axis, influences the transition weaker than $\mathrm B_{||} $ does and shifts the transition point, identified as the $C_m$ maximum, towards lower temperatures. However, it must be taken into account that in the PPMS system used, the calorimeter is suspended by 8 thin wires and in $\mathrm B_\bot $ a torque is applied to the sample. As the result, the sample tilts and the $c$ axis is no longer perpendicular to the applied field. Thus, we analyzed whether the shift of the transition can be attributed to tilting of the sample only, i.e., we assumed that the perpendicular to the $c$ axis component of the field has no influence on the transition and only the parallel component, appearing in the result of tilting, affects the transition. Then, based on the phase diagram constructed for $\mathrm B_{||}$, presented in Fig.~1f, we determined what $\mathrm B_{||} $, i.e.\ what tilting angle, would be necessary to cause the observed shift of the transition. As indicated in the legend of Fig.~\ref{fig4}, the expected tilt angles are quite probable and grow monotonically up to $20^\circ$. Thus, we find the assumption, that the transition is insensitive to $\mathrm B_{\bot}$, to be well-grounded.

\section{Conclusions}
Based on the detailed studies of specific heat, $C_B$, and magnetization, $M$, below 1~K, we found a magnetic phase transition in TbAl$_{3}$(BO$_{3}$)$_4$, which shifts to lower temperatures with increase of magnetic field $\mathbf B_{||}$, parallel to the easy magnetization axis. Determined behaviors of both $C_B$ and the Gr\"uneisen ratio, $\Gamma$, as a function of $T$ (especially scaling of the latter for $B_{||}\ge 0.30$~T), as well as dependence of $\Gamma$ on $B_{||}$ are characteristic of systems, in which the classical phase transition line is influenced by QF and ends at QCP. Vanishing of the nuclear specific heat for $B_{||}>0.4 75$~T confirms that $B_{||}$ destroys the long range magnetic order. The value of the dynamical critical exponent $z$ was assessed to $0.82 \le z \le 0.96$. However, a physical nature of the transition is not clear. The interpretation that this is the transition to the ferromagnetic order of Tb$^{3+}$ magnetic moments is the most natural, intuitive, and supported by the studies of $M$. However, such a classical transition should be smeared and shifted to higher $T$ by $B_{||}$, while we observe the opposite effect. We attribute this to QF, which dominate the behavior of the system and destroy the long range order, i.e., we suppose the transition to have quantum character. On the other hand, the behavior observed would be consistent with the behavior of the transition to an antiferromagnetic phase, but the studies of $M$, e.g., lack of a metamagnetic transition, linear dependence of $M$ on $B_{||}$, and reaching saturation magnetization nearly equal to the magnetic moment of free Tb$^{3+}$ ions in small $B_{||} \sim 0.7$~T, contradict this interpretation. Also the possibility, that the transition is not a magnetic one but related to any other kind of ordering, e.g., a complex multipolar ordering, and the ordering of the Tb$^{3+}$ moments is a ``side effect'' only, related to strong magneto-electric effect present in these materials, can not be ruled out. Generally, a physical mechanism proposed must predict not only the decrease of the transition point with increase in $B_{||}$ but the also the other unusual behaviors, like  $C_\mathrm {cr} \sim T^{y_0}$ and divergence of $\Gamma$, found in the present studies. To elucidate the physical mechanism responsible for the transition, detailed neutron diffraction studies are necessary but due to the low transition temperature and the presence of boron, very highly absorbing neutrons, such studies would be very difficult, though possible \cite{zhang}.

\begin{acknowledgments}
This work was supported partially by the National Science Centre, Poland, under project No. 2018/31/B/ST3/03289.
\end{acknowledgments}


%

\end{document}